\documentclass{svjour3}                     
\usepackage{graphicx}
\usepackage{natbib}
\bibpunct{(}{)}{;}{a}{}{,}
\usepackage{amsmath,amssymb}      

\newcommand{\be}{\begin{equation}}
\newcommand{\ee}{\end{equation}}
\newcommand{\beq}{\begin{eqnarray}}
\newcommand{\eeq}{\end{eqnarray}}

\newcommand{\muG}{\mu{\rm G}}
\newcommand{\kms}{{\rm km~s^{-1}}}
\journalname{SSRv}

\begin{document}

\title{Non-thermal processes in cosmological simulations}

\author{K. Dolag \and
        A.M. Bykov \and
        A. Diaferio}

\institute{K. Dolag \at Max-Planck-Institut f\"ur Astrophysik,
P.O. Box 1317, D-85741 Garching, Germany
\email{kdolag@mpa-garching.mpg.de} \and
           A.M. Bykov \at A.F. Ioffe Institute of Physics and Technology, St. Petersburg,
           194021, Russia \email{byk@astro.ioffe.ru} \and
           A. Diaferio \at Dipartimento di Fisica Generale ``Amedeo Avogadro'', Universit\`a
  degli Studi di Torino, Torino, Italy \at
  Istituto Nazionale di Fisica Nucleare (INFN), Sezione di Torino, Via P.
Giuria 1, I-10125, Torino, Italy
             }

\date{Received: 17 September 2007; Accepted: 7 November 2007 }

\maketitle

\begin{abstract}
Non-thermal components are key ingredients for understanding
clusters of galaxies. In the hierarchical model of structure
formation, shocks and large-scale turbulence are
unavoidable in the cluster formation processes. Understanding
the amplification and evolution of the magnetic field 
in galaxy clusters is necessary for modelling both the heat
transport and the dissipative processes in the hot
intra-cluster plasma. The acceleration, transport and
interactions of non-thermal energetic particles are essential for
modelling the observed emissions. Therefore, the inclusion of the
non-thermal components will be mandatory for
simulating accurately the global dynamical processes in clusters. In this
review, we  summarise the results obtained with the simulations of
the formation of galaxy clusters which address the issues of
shocks, magnetic field, cosmic ray particles and turbulence.
\end{abstract}



\section{Introduction}
\label{Introduction}

Within the intra-cluster medium (ICM), there are several processes 
which involve non-thermal components, such as the magnetic field and cosmic rays (CRs). 
At first glance, many of
these processes can be studied numerically in
simplified configurations where one can learn how the individual
processes work in detail. For example, there are 
investigations which consider magneto-hydro-dynamical (MHD) 
simulations of cloud-wind
interactions \citep[see][ and references
therein]{2000ApJ...543..775G} or simulate the rise of relic radio
bubbles \citep[see][ and references
herein]{2005ApJ...624..586J,2005MNRAS.357..242R}. Such work
usually focuses on the relevance of CRs, turbulence
and the local magnetic fields within and around these
bubbles. In this review, we will instead concentrate on simulations of
non-thermal phenomena, which aim at understanding the relevance
of these phenomena for galaxy cluster properties or at unveiling possible origins of
the non-thermal radiation. So far, firm evidence for the presence of
non-thermal emission (at radio wavelengths) and for the presence
of cosmologically relevant extended magnetic fields has been found
only in galaxy clusters; in filaments only weak limits have been
found so far. For a more detailed discussion see the
reviews on radio observations by \citet{rephaeli2008} - Chapter 5, this volume, and \citealt{ferrari2008} - Chapter 6, this volume.

\section{Possible origins of non-thermal components within the
large-scale structure (LSS)}\label{sec:origin}

The origin of magnetic fields and of CRs in galaxy clusters
is still under debate. The variety of possible contributors to
magnetic fields ranges from primordial fields, battery and dynamo
fields to all the classes of astrophysical objects which can
contribute with their ejecta. The latter possibility is supported
by the observation of metal enrichment of the 
ICM, which is due to the ejecta of stars in
galaxies, see 
\citealt{werner2008} - Chapter 16, this volume, and \citealt{schindler2008} - Chapter 17, this volume.
The magnetic fields produced by all these
contributors will be compressed and amplified by the process of
structure formation. The exact amount of amplification and
the resulting filling factor of the magnetic field will depend on where
and when the contributor is thought to be more efficient. 

CRs are produced at shocks, both in supernova remnants and in
cosmological accretion and merger shocks. Furthermore, active
galactic nuclei (AGN) and
radio galaxies contribute to the CR population.
CR protons can also produce CR electrons via
hadronic reactions with the thermal plasma, leading to the so-called secondary
CRs.

\subsection{Possible origins of magnetic fields}

Magnetic fields can be produced either at relatively low 
redshift ($z \sim 2-3$) or at high redshift ($z \gtrsim 4$). 
In the former case, galactic winds
\citep[e.g.][]{Volk&Atoyan..ApJ.2000} or AGN ejecta
\citep[e.g.][ and references therein]{1997ApJ...477..560E,Furlanetto&Loeb..ApJ2001} produce
magnetic fields  `locally', e.g. within
the proto-cluster region. In the latter case, 
the magnetic field seeds can also be 
produced by an early population of dwarf starburst galaxies or AGN
before galaxy clusters form gravitationally bound systems.

One of the main
arguments in favour of the 'low-redshift' models is that the high metallicity
observed in the ICM suggests that a significant enrichment occurred
in the past due to galactic winds or AGN. Winds and jets should carry
magnetic fields together with the processed matter. It has
been shown that winds from ordinary galaxies give rise to magnetic
fields which are far weaker than those observed in galaxy
clusters, whereas magnetic fields produced by the ejecta of starburst
galaxies can be as large as $0.1\,\muG$. Clearly, this class of
models predicts that magnetic fields are mainly concentrated in
and around galaxies and within galaxy clusters. Note that if the magnetic 
pollution happens early enough (around $z \sim 3$), these fields will be
amplified not only by the adiabatic compression of the proto-cluster
region, but also by shear flows, turbulent motions, and merging
events during the formation of the cluster. Shocks are
expected to be produced copiously during the non-linear stage of
the LSS formation (see below). Recent detailed studies of shock 
propagation revealed the presence of specific instabilities driven by
energetic accelerated particles
\citep{2001MNRAS.321..433B,2006ApJ...652.1246V}. Such instabilities
result in a strong, non-adiabatic amplification of an 
upstream magnetic field seed which converts an appreciable fraction of the
shock ram pressure into magnetic fields.

In the 'high-redshift' models of the magnetic field generation, 
the strength of the field seed
is expected to be considerably smaller than in the previous
scenario, but the adiabatic compression of the gas and the shear flows
driven by the accretion of structures can give rise to a
considerable amplification of the magnetic fields. Several
mechanisms have been proposed to explain the origin of magnetic
field seeds at high redshift. Some of them are similar to those
discussed above, differing only in the time when the magnetic
pollution is assumed to take place. In the present class of
models the magnetic field seeds are supposed to be expelled by an
early population of AGN or dwarf starburst galaxies at a
redshift between 4 and 6 \citep{Kronberg..1999ApJ}; this 
process would magnetise a large fraction of the volume. Recently, the
validity of such a scenario has been confirmed by a semi-analytic
modelling of galactic winds \citep{2006MNRAS.370..319B}.

Alternative models invoke processes that took place in the early
universe. Indeed, the presence of magnetic fields in almost all 
the regions of the universe suggests that they may have a cosmological 
origin. In general, most of the 
`high-$z$ models' predict magnetic field seeds filling the entire
volume of the universe. However, the assumed coherence length of
the field crucially depends on the details of the model. While
scenarios based on phase transitions give rise to coherence
lengths which are so small that the corresponding fields have
probably been dissipated, magnetic fields generated at neutrino or
photon decoupling are thought to have a much higher chance of 
surviving until the
present time. Another (speculative) possibility is that the 
field seed was produced during inflation. In this case, the coherence
length can be as large as the Hubble radius. See
\citet{Grasso..PhysRep.2000} for a review.

Magnetic field seeds can also be produced by the so-called
Biermann battery effect \citep{1997ApJ...480..481K,Ryu..1998}. The
idea here is that merger$/$accretion shocks related to the
hierarchical structure formation process give rise to small
thermionic electric currents which, in turn, may generate magnetic
fields.  The battery process has the attractive feature of being
independent of unknown physics at high redshift. Its drawback is
that, due to the large conductivity of the ICM, it can give rise
to very tiny magnetic fields, of order $10^{-21}$ G at most. One
therefore needs to invoke a subsequent turbulent dynamo to boost
the field strength to the observed level. 

Such a turbulent amplification, however, cannot be simulated numerically yet,
making it quite difficult to predict how it would proceed in a
realistic environment. It is clear that one expects the level of
turbulence to be strongly dependent on the environment, and that
it should appear mostly in high-density regions like collapsed
objects. While energetic events, such as mergers of galaxy
clusters, can easily be effective at driving the required levels of
turbulence, it is harder to understand how turbulent 
amplification would work in relatively quiet
regions like filaments. Here, magnetic fields can be rather
amplified by the shocks which originate during the LSS formation,
as mentioned at the beginning of this section. Lacking a
theoretical understanding of the turbulent amplification, it is
therefore not straightforward to relate the very weak field seeds
produced by the battery process to the magnetic fields observed
today. Attempts to construct such models, based on combining
numerical and analytical computations, have not yet been reported
to successfully reproduce the observed scaling relations of the
magnetic fields in galaxy clusters.

\subsection{Possible origins of CRs}

There are many possible contributors to the CR population
in the ICM. AGN and stellar activity in cluster
galaxies are believed to be a significant source of CRs 
in the ICM. Furthermore, cluster accretion shocks of high Mach
numbers (typically above 100) and of sizes of megaparsec or
larger are widely accepted to be possible CR sources with
nuclei accelerated up to $\sim 10^9$ GeV
\citep[e.g.][]{1995ApJ...454...60N}. Cluster mergers generate
internal shocks (of moderate Mach numbers, typically below 4)
which provide most of the ICM gas heating \citep[see
e.g.][]{2005ApJ...620...21K}, but are also likely to convert a
non-negligible fraction ($\gtrsim$ 10~\%) of their power into CRs.

AGN outflows dissipate their kinetic energy and Poynting fluxes into
the ICM  providing additional non-gravitational ICM heating and,
thus, a plausible feedback solution to the cooling flow problem
\citep[e.g.][]{1995MNRAS.276..663B,2003ApJ...590..225C}. Recent
{\sl XMM-Newton} observations of the cluster MS 0735+7421
\citep{2007ApJ...660.1118G} revealed that a powerful AGN outburst 
deposited about 6$\times$10$^{61}$ erg into the ICM 
outside the cooling region of radius $\sim$ 100 kpc. Relativistic
outflows of AGN are likely to be an essential source of
super-thermal particles in clusters. A powerful relativistic AGN
jet could deposit up to 10$^{62}$ erg into a relativistic particle
pool during a duty cycle of about $\sim$ 50 Myr. A potential
source of energetic particles are sub-relativistic ions with
energy $\sim$ 100 MeV that are evaporated from an AGN, which accretes
mass in the ion-supported tori regime
\citep[e.g.][]{1982Natur.295...17R}. A certain fraction of the
energetic particles could escape the flows and avoid fast
cooling  due to collective effects in the central parts of the
cluster. Re-acceleration of that non-thermal population by inner
shocks inside the cluster can provide a long-lived non-thermal
component that contributes to the total pressure of the ICM.

\begin{figure}    
\begin{center}
\includegraphics[width=0.5\textwidth]{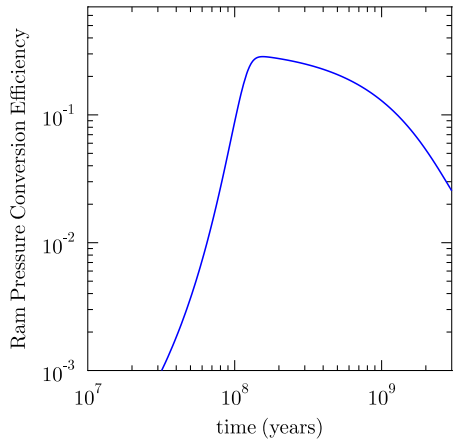}
\end{center}
\caption{The simulated efficiency of conversion of shock and bulk
MHD motions ram pressure power to the CRs at  different stages
of cluster evolution. The efficiency is defined as a volume
averaged fraction of the cluster large-scale motions power
transferred to CRs. The simulation was done for a model of CR
acceleration by an ensemble of shocks and MHD-turbulent motions
described in \citet{2001SSRv...99..317B}, adopted for clusters of galaxies. The
MHD turbulence in a cluster of a Mpc size was initiated  with the
Gaussian bulk velocity dispersion of 2\,000~$\kms$ at a scale of 100
kpc.}\label{n-t-power}
\end{figure}

Star formation activity in galaxies is another source of CRs in
clusters of galaxies \citep[e.g.][ and references therein]{1996SSRv...75..279V}. The combined action of supernovae and winds of 
early-type stars leads to the formation of a hot, X-ray emitting,
slowly expanding superbubble filled with large-scale (tens of
parsecs) compressible MHD motions (e.g. rarefactions) and shocks.
When the starburst event is energetic enough, the superbubble may
expand beyond the disk of the parent galaxy and produce a superwind
that supplies the intergalactic medium with metals ejected by
supernovae and CRs. The superwinds have been seen with Chandra in
some starburst galaxies: M~82, Arp 220  and NGC 253. In the
superwind scenario, the CR ejection is tightly connected with the
metal enrichment of the cluster (\citealt{schindler2008} - Chapter 17, this
volume). 

\citet{2001SSRv...99..317B}
argues that non-thermal particle acceleration
can favourably occur in correlated supernovae events and powerful stellar
winds with great energy release, which generate 
interacting shock waves within superbubbles.  The acceleration
mechanism provides efficient creation of a non-thermal nuclei
population with a hard low-energy spectrum; this population can transport a
substantial fraction ($\sim$ 30\%) of the kinetic energy released
by the supernovae and by the winds of young massive stars. 

A bright phase
in the galaxy evolution can be the source of the relic CRs 
in clusters. Energetic nuclei can be stored in cluster magnetic
fields for several Hubble times
\citep[e.g.][]{1997ApJ...487..529B}. The presence of these nuclei
produces a diffuse flux of high-energy $\gamma$ and neutrino
radiation due to the interaction of the CRs with the
ICM \citep{1996SSRv...75..279V,1997ApJ...477..560E}. The resulting flux depends on the ICM baryon density.

A physical model of particle acceleration by the ensemble of inner
shocks in the ICM could be similar to the
superbubble model. The sub-cluster merging processes
and the supersonic motions of dark matter halos in the ICM are accompanied
by the formation of shocks, large-scale flows and broad spectra of
MHD-fluctuations in a tenuous intra-cluster plasma with frozen-in
magnetic fields. Vortex electric fields generated by the 
large-scale motions of highly conductive plasma harbouring shocks result in a
non-thermal distribution of the charged nuclei. The free energy
available for the acceleration of energetic particles is in the ram
pressure of the shocks and in the large-scale motions. 

The most studied way to transfer the power of the MHD motions to the
energetic particle population is the Fermi-type acceleration
(see e.g. the review by \citealt{1987PhR...154....1B} and \citealt{bykov2008} - Chapter 7, this volume). An important ingredient of 
the energetic particle acceleration by shocks and
large-scale MHD motions is the presence of small-scale MHD
turbulence, which is necessary to scatter relativistic particles and to
make their pitch-angle isotropic. The scale of the fluctuations
required for the resonant scattering of a particle of energy
$\sim$ 1 GeV is about  3$\times$ 10$^{12} B^{-1}_{-6}$ cm, where
$B_{-6}$ is the local mean magnetic field in $\mu$G. The scale is
some 10--11 orders of magnitude smaller than the basic energy
scale of the system; thus the origin and the maintenance
mechanisms of such small scale turbulence is a serious problem. 

In non-linear models of particle acceleration by strong MHD shocks
the presence of turbulence could be supported by the CR
instabilities themselves \citep[see e.g. recent non-linear
Monte-Carlo simulations by][]{2006ApJ...652.1246V}. Direct MHD
cascade of energy to small scale turbulence is also possible, but
the small scale fluctuations are highly anisotropic in that case
\citep[see e.g.][]{2003matu.book.....B}. Moreover, the cascade
properties are still poorly known at the scales close to the
Coulomb mean free path where the transition from the collisional
to the collisionless plasma regime occurs.

\begin{figure}    
\begin{center}
\includegraphics[width=0.5\textwidth]{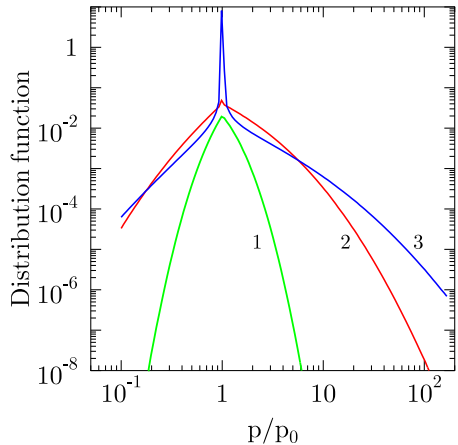}
\end{center}
\caption{The temporal evolution of the particle distribution function
in a cluster of a Mpc size and a Gaussian bulk velocity
dispersion of about 2\,000~$\kms$ at a scale of 100 kpc. The curves
labelled as 1, 2, 3 correspond to 0.1, 0.3 and 1 Gyr age of the
cluster. Mono-energetic proton injection at energy 10 keV ($p_0$
is the injection momentum). The injection efficiency (see Fig.~\ref{n-t-power}) was assumed
to be about 10$^{-3}$. The power conversion efficiency for the
model was already illustrated in Fig.~\ref{n-t-power}.}\label{n-t-spectra}
\end{figure}

The particle distribution within a system with multiple shocks and
with large-scale plasma motions is highly intermittent, because
it has strong peaks at shocks and a non-steady mean CR background. 
By using the kinetic equations given by
\citet{2001SSRv...99..317B}, one can construct a model of
the temporal evolution of the particle distribution function, which
accounts for the non-linear effect of the reaction of the
accelerated particles on the shock turbulence inside the cluster.
In Fig.~\ref{n-t-power} we show the
efficiency of a conversion into CRs of the power of shocks and the
bulk plasma motions of a scale of about 100 kpc. One may note
that, while the efficiency could be as high as 30~\% for some
relatively short period of time, the efficiency is typically $\sim$ 10~\% for
most of the time. Fig.~\ref{n-t-spectra} shows the temporal behaviour
of the particle spectra on a Gyr time scale. 

Non-thermal particles and magnetic fields could contribute to the
total pressure of the ICM. In the particular model described above
a substantial energy density could be stored in protons of energy
below 200 MeV. Protons of that energy level produce very little
high energy emission and therefore it is rather difficult to
constrain that component directly from the 
observational data which are currently available.

\section{Shocks in the LSS and clusters}

As described in the previous section, shocks can play a crucial role as
source of non-thermal components in galaxy clusters. Therefore,
a significant effort is spent to investigate the distribution and
the evolution of shocks during structure formation. 

In early simulations of the formation of cosmological
structures, it was already
noticed that shocks are the main source of the ICM heating
\citep[see][ and references therein]{1998ApJ...502..518Q}. In
numerical simulations, two classes of shocks can be identified.

The first one is the accretion shocks surrounding the forming
objects, namely filaments or galaxy clusters. In the case of galaxy
clusters, most shocks are quasi-spherical and their position with
respect to the centre of the cluster grows with time (shock
travelling outwards). These shocks can be associated with the accretion of
diffuse matter which is relatively cold. These shocks provide a first step towards
the virialisation of an initially supersonic accreting flow. The
radius of the accretion shock (for a quasi-spherical cluster at
$z=0$) is typically two times the virial radius. As the typical
temperature and density of the upstream gas is quite low, such
shocks can have very large Mach numbers ($\gg10$).

The second class of shocks in galaxy clusters arises from the
merging of substructures. The gas in gravitationally bound
substructures is much denser than the diffuse ICM; therefore it cannot
be stopped in the accretion shock and it continues to move with the
substructure through the gas, which is mostly virialised. The substructures,
which have velocities larger than the sound velocity of the hot ICM,
drive the so-called merger shocks. These shocks are initiated
within the ICM which has been already shock-heated, and therefore 
they typically have
modest Mach numbers (below $\approx$4). The merger shocks propagate
within the dense ICM and are likely to be the main
contributor to the global energy dissipation within the ICM
\citep[see][ and references therein]{2006MNRAS.367..113P}. As these
shocks are primarily driven by gravitational attraction,
non-gravitational processes (like cooling or feedback by star
formation processes) do not significantly affect the global energy
dissipation \citep[see][]{2007arXiv0704.1521K}. The ram pressure
dissipation process in shocks within the collisionless plasma is
non-trivial because of the role of long-lived non-thermal
components, namely magnetic fields, MHD-waves and energetic particles
(see for details \citealt{bykov2008} - Chapter 7,  this volume).
The complex multi-scale structure of a shock modified by energetic CRs 
(e.g. where the back reaction of the CRs on the thermal plasma is
strong enough to change the thermodynamical states of the plasma
in addition to the shock itself)
cannot be resolved properly in LSS simulations.

\begin{figure}
\begin{center}
\includegraphics[width=0.5\textwidth]{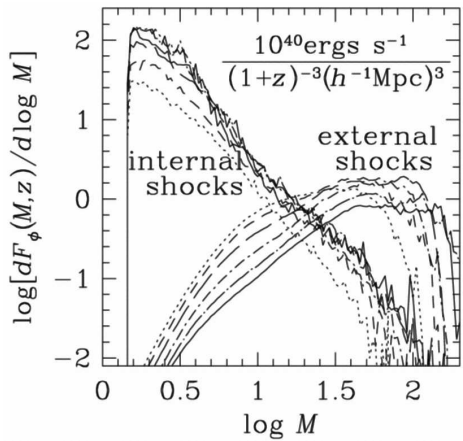}
\end{center}
\caption{Shocks detected in a cosmological simulation of a cubic
volume with a side length of 100~Mpc,
resolved with 512$^3$ grid points. The figure shows the kinetic energy flux per unit 
comoving volume through surfaces of external and internal shocks 
with Mach number between ${\rm log} M$ and 
${\rm log} M + {\rm d}({\rm log} M)$ 
at different redshifts (solid, $z=0$ to
dotted, $z=2$). From \citet{2003ApJ...593..599R}. }
\label{fig:Ryu}
\end{figure}

The efficiency of energising CRs by shocks strongly depends on a
number of factors. Among the most important factors, we have
(i) the upstream magnetic field of the shock, (ii)  the plasma
parameter $\beta = {\cal M}^2_{\rm a}/{\cal M}^2_{\rm s}$ (ratio
of thermal to magnetic pressure, here expressed by the Mach
numbers ${\cal M}_{\rm a}$ and ${\cal M}_{\rm s}$ with respect to the Alfv\'en and sound
velocity respectively), (iii) the inclination of the upstream field of the shock,  and
(iv) the presence of a pre-existing CR population produced by the
accretion shock or in star-forming regions.

The distribution of the Mach numbers of the shocks
within the cosmological structure can be derived from
semi-analytical modelling \citep[see][ and references
therein]{2003ApJ...583..695G,2006ApJ...642..734P}, as well as
directly from cosmological simulations \citep[see][ and references
therein]{2000ApJ...542..608M,2003ApJ...593..599R,
2006MNRAS.367..113P,2007arXiv0704.1521K}. An example is shown in
Fig.~\ref{fig:Ryu}. There are significant differences in the strength
and the distribution of the shocks found in different works: this fact has
strong consequences on the predicted amount and the relative
importance of CRs and their energy with respect to the thermal
energy of the ICM. Note that part of the confusion, specially in
the early work, is caused by a lack of resolution within the
simulations, which prevents internal shocks from being properly resolved 
\citep[see][]{2003ApJ...593..599R,2006MNRAS.367..113P}; further
confusion originates from the influence of other, non-thermal
processes (like preheating) on the inferred Mach number of
the external shocks
\citep[see][]{2006MNRAS.367..113P,2007arXiv0704.1521K}. Moreover,
differences in the underlying hydrodynamic simulations as well as in the
shock detection, mostly done in a post processing fashion, can
contribute to differences in the results. Nevertheless, there is a
qualitative agreement in the case of simulated internal shocks between
the simulations done by \citet{2003ApJ...593..599R} and
\citet{2006MNRAS.367..113P}.

\section{Evolution of magnetic fields in simulations}

\subsection{Local amplification of magnetic fields}\label{subsec:localampl}

\begin{figure}
\begin{center}
\includegraphics[width=0.85\textwidth]{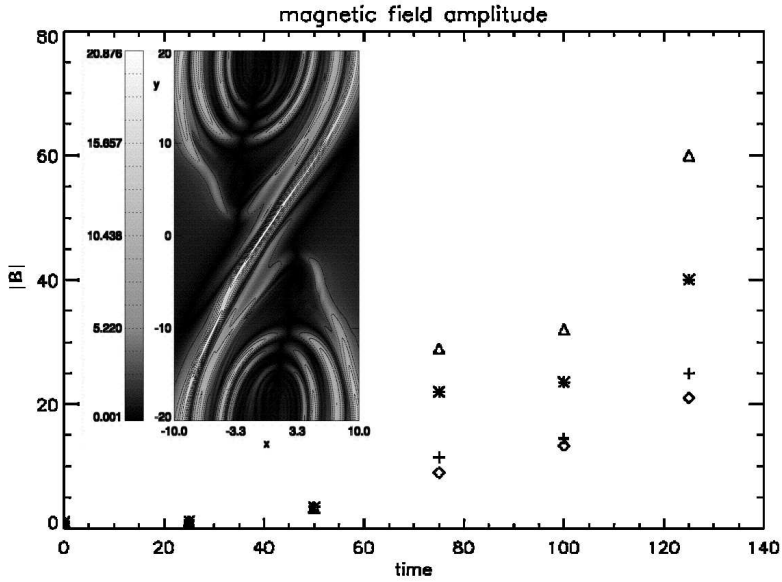}
\end{center}
\caption{Resistive, MHD simulation of a flow which is KH instable
at the interface separating the left and right halve of the simulated 
volume. Shown is the magnetic field (y-component) in units of the initial magnetic field after evolving the KH 
instability (inlay) and the time evolution of the (relative) 
magnetic field amplitude for different values of the 
chosen resistivity and initial magnetic field. Time is in units of the Alfv\'enic transit time (4000 year). Triangles and stars are for an initital magnetic field of 10~$\muG$, crosses and diamonds for a ten times larger field. Stars and diamonds have a magnetic Reynolds number of 100, triangles and crosses have a ten times larger magnetic Reynolds number.
From \citet{1999ApJ...518..177B}. } \label{fig:Birk}
\end{figure}

A very basic process which amplifies magnetic fields is
related to the Kelvin-Helmholtz (KH) instabilities driven by shear
flows, which are common during the formation of cosmic structures.
\citet{1999ApJ...518..177B} performed a detailed study
of such an amplification within the outflows of starburst galaxies, where the
KH timescale should be $\approx 4\times10^5$ years (see Fig.~\ref{fig:Birk}). By using a
Cartesian resistive MHD code they found that the 
amplification factor of the magnetic field mainly depends on the
initial ratio of the magnetic to the kinetic energy and only mildly
depends on the assumed resistivity. They concluded 
that such a process could indeed explain why 
the magnetic field observed in the halo of starburst galaxies is
significantly higher than 
what is expected from the magnetic fields observed within galactic
disks. When applied to a cluster core environment, the KH
timescale turns out to be $10^7$ years; this timescale makes 
the KH instability an interesting
process to amplify weak magnetic fields. 

\begin{figure}
\begin{center}
\includegraphics[width=0.85\textwidth]{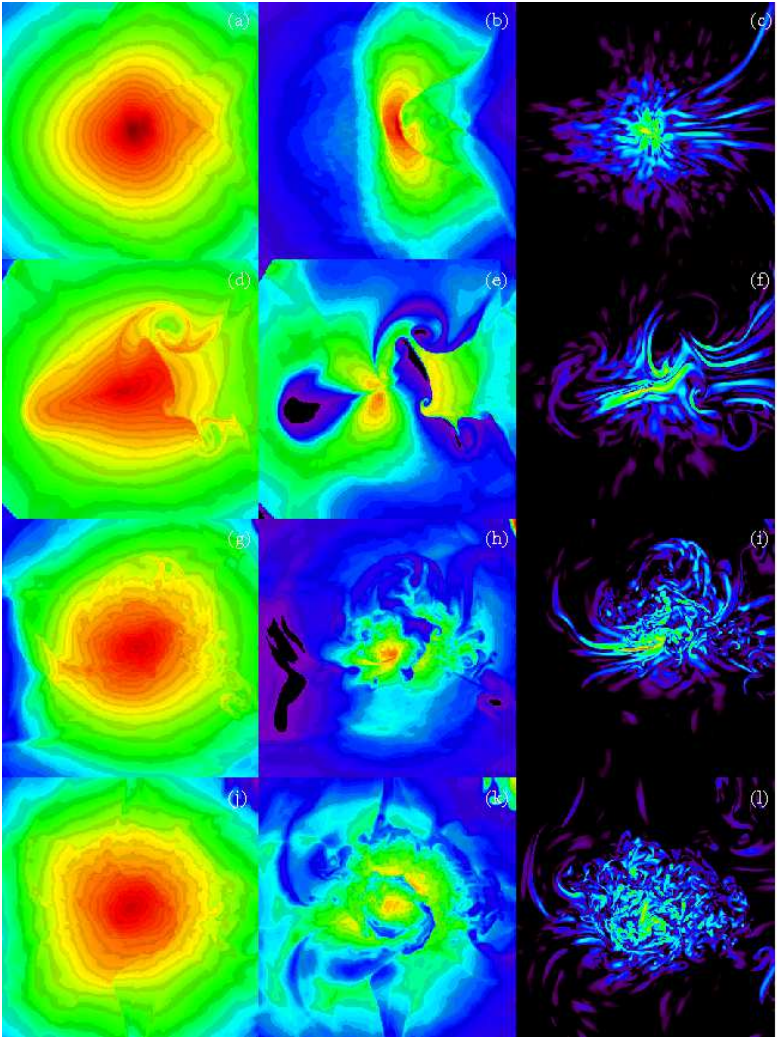}
\end{center}
\caption{The evolution of (the logarithm of) the gas density (left
column), the gas temperature (central column), and (the logarithm of)
the magnetic pressure (right column) in two-dimensional slices taken
through the core of a cluster in a major merger phase in the plane 
of the merger.  Each row
refers to different epochs: $t=0$ (i.e. the time of the core
coincidence), $t=1.3$, $t=3.4$, and $t=5.0$ Gyr, from top to
bottom. Each panel is $3.75 \times 3.75$ Mpc. The merging subcluster enters from the right. From
\citet{1999ApJ...518..594R}. } \label{fig:roet_I_a}
\end{figure}

It is quite
expected that merger events and accretion of material onto galaxy
clusters will drive significant shear flows within the ICM.
Extensive MHD simulations of single merging events performed using
the Eulerian code {\it ZEUS}
\citep{1992ApJS...80..753S,1992ApJS...80..791S} demonstrated 
that merger events are effective at amplifying magnetic fields 
\citep{1999ApJ...518..594R}.  In particular these authors found that 
the field initially becomes quite filamentary, as a result of stretching and
compression caused by shocks and bulk flows during infall; at
this stage 
only a minimal amplification occurs. When the bulk flow
is replaced by turbulent motions (e.g., eddies), the field amplification
is more rapid, particularly in localized regions, see Fig.~\ref{fig:roet_I_a}. The total magnetic field energy is found to
increase by nearly a factor of three with respect to a non-merging
cluster. In localised regions (associated with high vorticity),
the magnetic energy can increase by a factor of 20 or more. 

A power spectrum analysis of the magnetic energy showed that the
amplification is largely confined to scales comparable to or
smaller than the cluster cores: this indicates that the core
dimensions define the injection scale.  It is worth taking notice
that the previous results can be
considered as a lower limit on the total amplification, because of 
the lack of resolution. Furthermore,
it is quite likely that a galaxy cluster undergoes more than one
of these events during its formation process, and that the
accretion of smaller haloes also injects turbulent motions into the
ICM: consequently, the magnetic field amplification within galaxy
clusters will be even larger. A detailed discussion of the
amplification of magnetic field in the cluster environment, using
various simulations of driven turbulence, can be found in
\citet{2006MNRAS.366.1437S}; these authors show that 
turbulent processes can provide reasonable strength
and length scales of the magnetic fields in galaxy clusters.

\subsection{Magnetic fields from cosmological shocks}

\begin{figure}
\begin{center}
\includegraphics[width=0.85\textwidth]{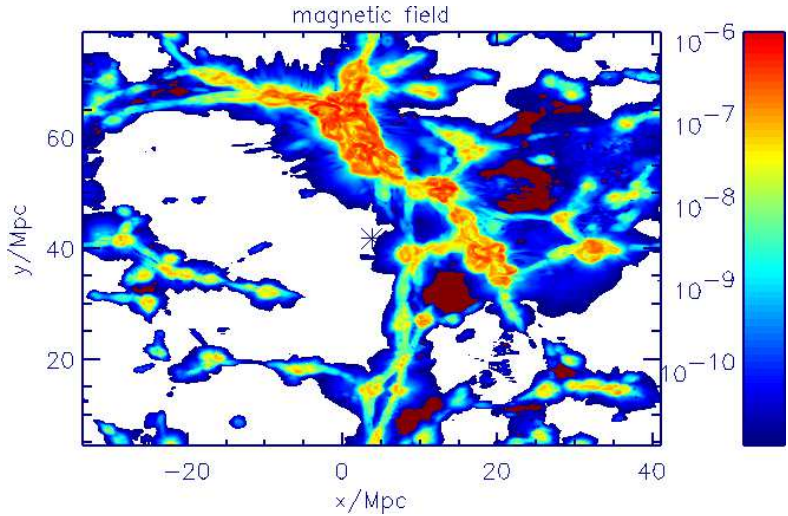}
\end{center}
\caption{A two-dimensional cut through a cosmological box
simulated by \citet{2001ApJ...562..233M}, following the
evolution of battery fields. The logarithm of the
up-scaled magnetic field strength in Gauss at $z=0$ is shown. Taken
from \citet{2004PhRvD..70d3007S}. } \label{fig:Sigl_I}
\end{figure}

Cosmological shocks, mainly the accretion shocks on cosmological
objects like galaxy clusters and filaments, occur more frequently
than the ones produced by individual merger events. They also can 
produce magnetic fields through the so-called Biermann battery effect
\citep{1997ApJ...480..481K,Ryu..1998}, on which a subsequent
turbulent dynamo may operate (see Sect.~\ref{sec:origin}). 

In such a scenario, the magnetic field is strongly correlated with the
large-scale structure (see Fig.~\ref{fig:Sigl_I}). This means that
the magnetic field within the filamentary structure could be even
slightly higher than its equipartition value without violating the
(weak) upper limits of the rotation measure (RM) of quasars, as pointed out
by \citet{Ryu..1998}. It is worth pointing out that the arguments
for the turbulent dynamo action, which could amplify the battery
field seeds up to the $\mu$G level, as presented in
\citet{1997ApJ...480..481K}, refer explicitly to regions which
are close to collapse into galaxies. It has still to be proven that such
arguments hold within proto-clusters or even cosmological
structures, like sheets and filaments. In general, the time
evolution of the magnetic field, as predicted by these simulations,
strongly flattens around $z\approx3$ \citep{1997ApJ...480..481K};
the evolution then leads to a magnetic field intensity which is relatively uniform on
scales of tens of Mpc within the LSS around galaxy clusters (see
Fig.~\ref{fig:Sigl_I}). Note that, so far, the
synthetic rotation measures due to the magnetic fields predicted
by up-scaling the battery fields have never been compared with 
the rotation measure observed on the scale of galaxy clusters. 
This might be partially motivated by the lack of
resolution in the mock clusters within such non-adaptive,
Eulerian simulations.

\subsection{Cosmological MHD simulations}

\begin{figure}
\begin{center}
\includegraphics[width=0.85\textwidth]{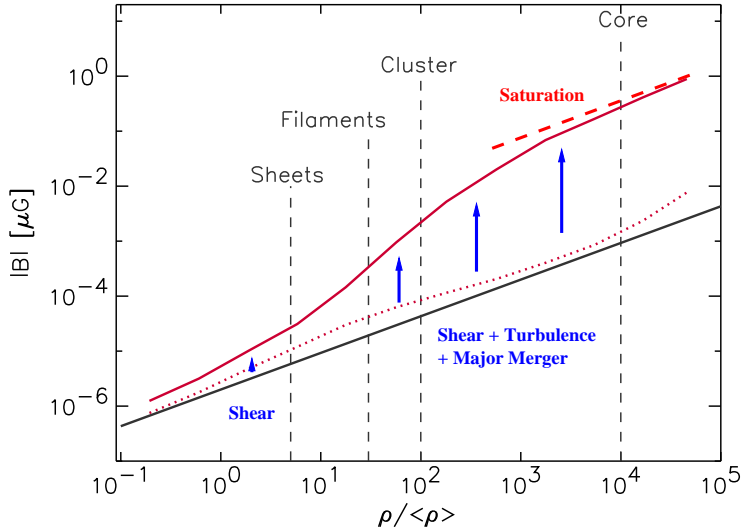}
\end{center}
\caption{The strength of the magnetic field  as a function of baryonic
overdensity within a cosmological simulation. 
The straight line shows the expectation for a purely
adiabatic evolution, the solid line gives the mean field strength
at a given overdensity within a cosmological simulation
\citep{2005JCAP...01..009D}. While the latter is close to the
adiabatic value in underdense regions, in clusters there is a significant
inductive amplification due to shear flows and
turbulence; this amplification however saturates in the cluster cores. At
any given density, a large fraction of particles remains close to
the adiabatic expectation, as shown by the dotted line, which
gives the median of the distribution at each density.}
\label{fig:b_ampli}
\end{figure}

By using the {\it GrapeMSPH} code \citep{1999A&A...348..351D} and assuming
that a small initial magnetic field seed exists before structure
formation (see Sect.~\ref{sec:origin}), the first
self-consistent simulations which follow the magnetic field
amplification during the formation of galaxy clusters within a
cosmological environment have been performed by
\citet{1999A&A...348..351D,2002A&A...387..383D}. These simulations
demonstrated that the contribution to the amplification of
magnetic fields by shear flows (and by its induced turbulence) is
significant (see Fig.~\ref{fig:b_ampli}). Therefore, for the first
time, a consistent picture of the magnetic field in galaxy
clusters could be constructed: the amplification predicted by the
simulations was capable of linking the predicted strength of the 
magnetic field seed (see Sect.~\ref{sec:origin} and references
therein) at high redshift ($z\approx3$ and higher) to the 
observed magnetic field strength in galaxy clusters in 
the local universe.

\begin{figure}
\begin{center}
\includegraphics[width=0.8\textwidth]{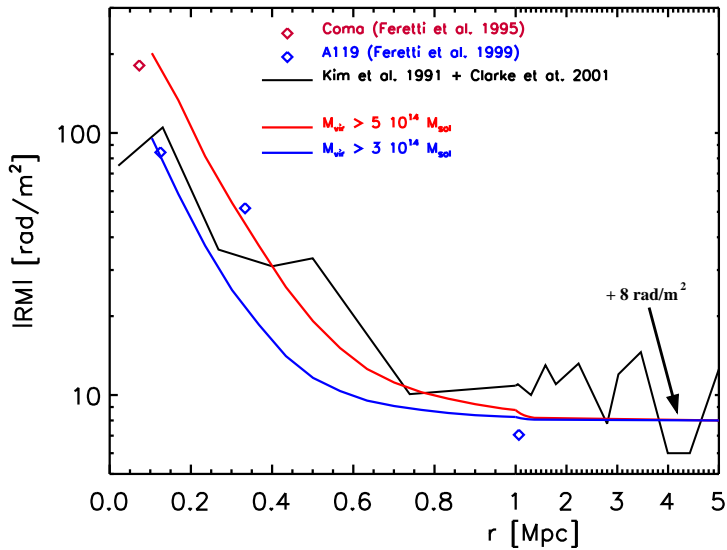}
\end{center}
\caption{Comparison of RMs in simulations with observations
of Abell clusters, as a function of the distance from the cluster centre. 
The smooth lines represent the median values of $|{\rm
RM}|$ produced by simulated clusters with masses above $5
\times 10^{14}~{\rm M}_\odot$ (upper line) and $3 \times 10^{14}~{\rm
M}_\odot$ (lower line). The broken line represents the median of combined data
taken from the independent samples presented in
\citet{1991ApJ...379...80K} and \citet{2001ApJ...547L.111C}. We
also include data (diamonds) for the three elongated sources
observed in A119 \citep{1999A&A...344..472F}, and for the
elongated source observed in the Coma cluster
\citep{1995A&A...302..680F}.} \label{fig:RMprof}
\end{figure}

Furthermore, the simulations predicted that the final structure of
the magnetic field in galaxy clusters reflects the process of
structure formation, and no memory on the initial magnetic field
configuration survives: this relaxes the constraints on models for
magnetic field seeds. In general, such models predict a magnetic
field profile similar to the density profile. Thereby the
predicted rotation measure (RM) profile agrees with the observed
one (see Fig.~\ref{fig:RMprof}). \citet{Dolag:2001} found  a quasi 
linear correlation between two observables, namely the X-ray
surface brightness and the RM r.m.s., both in observations and
in their MHD simulations.
By extending the {\it GADGET2} code to follow the full set of ideal MHD
equations, \citet{2004JETPL..79..583D,2005JCAP...01..009D}
performed several realizations of a cosmological volume which confirm
these results at even much higher resolution.

\begin{figure}
\begin{center}
\includegraphics[width=\textwidth]{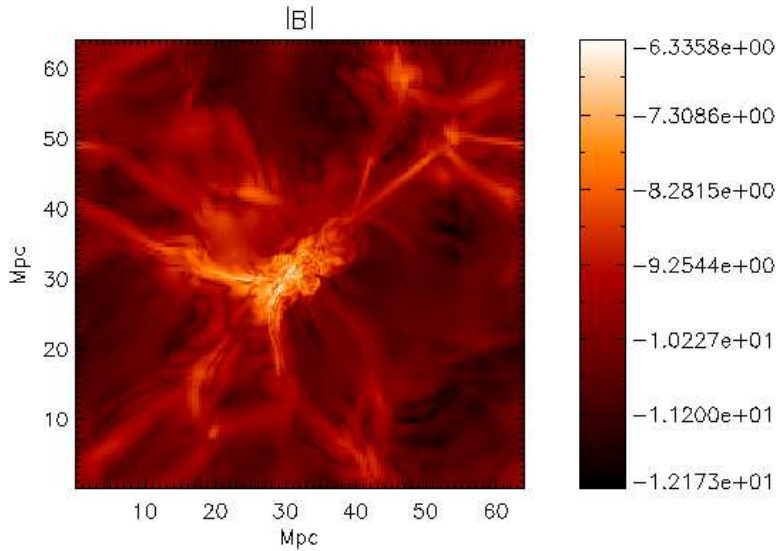}
\end{center}
\caption{ Slice through the centre of the simulated box hosting a
massive galaxy cluster. The simulation used the AMR code {\it
FLASH} and followed the evolution of a weak magnetic field seed.
The figure shows the logarithm of the magnetic field strength, in Gauss, 
measured at $z=0.5$. From \citet{2005ApJ...631L..21B}.}
\label{fig:Brueggen_I}
\end{figure}

Recently, \citet{2005ApJ...631L..21B} confirmed all the previous results,
which are based on SPH codes, with a simulation of
the formation of a single galaxy cluster in a cosmological
framework, using a passive MHD solver implemented into FLASH,
an adaptive mesh refinement (AMR) code (see Fig.~\ref{fig:Brueggen_I}).

Another interesting quantity to look at is the slope $\alpha$ of
the magnetic field power spectrum ($\propto k^{-\alpha}$, with $k$
being the wave vector). Within galaxy clusters, $\alpha$ is
predicted by the SPH simulations
\citep{2002A&A...387..383D,2004APh....22..167R} to be slightly
lower, but still very close to $11/3$, which is the expected value
for a Kolmogorov-like spectrum in 3D. The AMR simulation by
\citet{2005ApJ...631L..21B} nearly perfectly matches the
Kolmogorov slope.

\subsection{Approximative simulations}

The fully consistent cosmological MHD simulations described above are very
expensive in their requirements of computing resources; thus, for
the investigation of various models, simplified approaches are often
used in astrophysics. Recently, \citet{2006MNRAS.370..319B} used
semi-analytic simulations of magnetised galactic winds coupled to
high-resolution N-body simulations of structure formation to
estimate limits for the fraction of the ICM which can be
significantly magnetised. Interestingly, the fraction of the
volume of the ICM, which is predicted to be magnetised at a
significant level by the galactic outflows during structure
formation, is very high. In fact, the volume filling is high enough
to be compatible with observations and a moderately efficient
turbulent dynamo, operating within the ICM, will easily amplify
these fields up to the observed values.

Additional to any dynamo action, further support for strong
magnetic field amplification during the process of structure
formation comes from the application of the Zel'dovich
approximation \citep{1970A&A.....5...84Z} to follow the MHD
equations during the gravitational collapse
\citep{2003MNRAS.338..785B,2006MNRAS.365.1288K}. These papers
demonstrate the existence of a super-adiabatic amplification 
due to the anisotropy
of the collapse of the LSS within the cold dark matter paradigm.

A novel aspect of including magnetic field pressure into LSS
simulations - even if in a simplified way - is investigated by
\citet{2007MNRAS.375..657G}, which followed a purely empirical
approach. Inspired by the scaling between magnetic field and
gas density, found in both simulations and observations \citep{Dolag:2001}, 
the magnetic field is parameterised as
$B(z)=B_0(1+z)^{3\alpha}\rho^\alpha$. Then an isotropic pressure
($\propto B^2$) term is added into the equation of motion.
This approach allows a quite effective way to identify 
which models of magnetic fields, along with their parameters, 
provide a non-thermal pressure support which is able to modify significantly the
formation of structures. Thereby this approach gives a unique insight
for the possible consequences of certain magnetic field scenarios.

\section{Cosmic Rays}\label{sec:CRs}

\subsection{Modelling of Cosmic Rays (CRs)}

CRs may originate in a number of cosmological processes
including jets due to mass accretion, large-scale shocks and
supernova remnants (see Sect.~\ref{sec:origin}). One can
conventionally distinguish two important
classes of CR modelling in cosmological simulations. 

The first one
is related to the direct modelling of the origin, spatial distribution and
spectral evolution of energetic electrons and positrons, which are
responsible for most of the non-thermal emission observed so far
in cosmological objects. In the vast majority of the non-thermal
lepton (${\mathrm e}^{\pm}$) simulations, ${\mathrm e}^{\pm}$ are considered as test
particles scattered and accelerated by electro-magnetic fields
associated with the cosmological flows. In the highly relativistic energy
regime, the leptons radiate very efficiently and therefore suffer
from radiative losses. Most of the models are based on the linear
kinetic equations for the electron distribution function without
accounting for the back-reaction that the accelerated electrons 
have on the bulk plasma motions and on the MHD-turbulence.

The second class of the CR simulations is related to the acceleration
and propagation of energetic baryons. The radiation inferred from the energetic
trans-relativistic baryons is basically caused by their nuclear
interactions with the ambient matter which produces $\gamma$-rays, neutrino
and ${\mathrm e}^{\pm}$-pairs. Non-relativistic super-thermal baryons can
produce some emission by inverse-Bremsstrahlung mechanisms. The
most important feature of that class of processes, however, is
that energetic non-thermal baryons could absorb a substantial
fraction of the ram pressure of both non-relativistic and
relativistic plasma flows. The back-reaction effects of energetic 
non-thermal particles on the plasma flows must then be taken
into account. The most popular example of such a non-linear
process is the diffusive particle acceleration by astrophysical
shocks \citep[see e.g.][ for a review]{1987PhR...154....1B,
2001RPPh...64..429M}. The process of diffusive particle
acceleration by MHD shocks could be very efficient in highly
turbulent cosmic plasma. 

In Fig.~\ref{n-t-power} of
Sect.~\ref{sec:origin} we illustrated the
simulated efficiency of the conversion into CRs of the power of shocks and
of the bulk plasma motions in the particular cosmological context which
represents the typical environment within galaxy clusters. 
The non-linear effect of the
back-reaction of accelerated particles on large-scale plasma flows
results in the specific temporal evolution of the particle spectra
shown in Fig.~\ref{n-t-spectra} of Sect.~\ref{sec:origin}.
Moreover, as it was mentioned above, under a certain condition, 
thanks to the conversion of a fraction of
the shock ram pressure into magnetic field energy, an
efficient acceleration of baryons by MHD shocks in a turbulent
cosmic plasma results in a strong amplification of the magnetic
field  in the shock upstream.

\begin{figure}
\begin{center}
\includegraphics[width=0.85\textwidth]{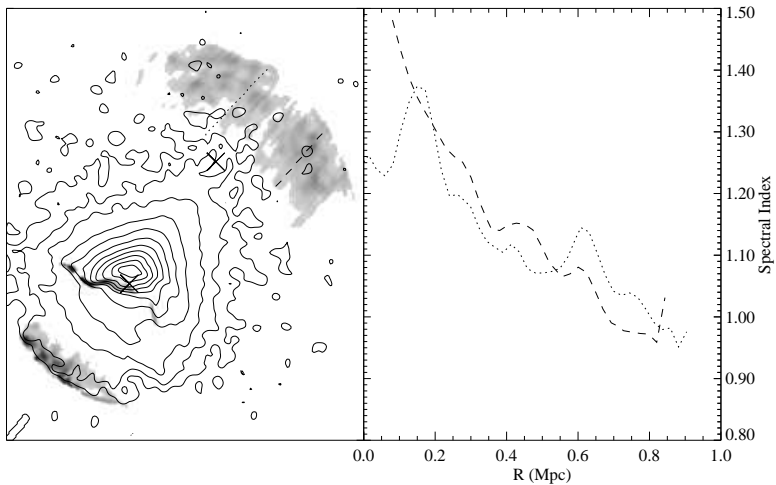}
\end{center}
\caption{Results from a simulation intended to
reproduce the merging system observed in A~3667. The left panel 
shows the simulated X-ray surface brightness 
(contours) and the radio emission at 1.4 GHz (grey scale).
The image is $3.15 \times 3.85$ Mpc. The dashed
and dotted lines refer to the location of the radio spectral index
($\alpha^{1.4}_{4.9}$) profiles displayed in the right panel.
From \citet{1999ApJ...518..603R}.} \label{fig:roet_II_a}
\end{figure}

The acceleration of relativistic electrons and their emission were modelled in
different cosmological contexts. To study the gradients of the spectral indices
inferred from the radio observations at $1.4-5.0$ GHz of five luminous 3C radio
galaxies, \citet{1985ApJ...291...52M} developed a model in which
the radiation is due to an isotropic ensemble of relativistic
electrons which are subject
to synchrotron radiation losses (synchrotron aging). \citet{1985ApJ...291...52M} inferred
the age of the electrons, namely the time since their acceleration, at various
locations in the lobes, and consequently derived the speed of the separation of
the hot spots from the lobe material. The inferred speeds were found to
be in the range $10\,000-30\,000$ km~s$^{-1}$. These results are
consistent with the beam model, in which the lobe material is left
behind by a hot spot advancing through the intergalactic medium at
speeds of about 10\,000 km~s$^{-1}$.

\citet{2001ASPC..250..454E} modelled different scenarios 
for the evolution of the radio plasma inside the
cocoons of radio galaxies, after the activity of the central
engine has ceased.
These authors discussed an analytical model for the evolution of a
relativistic electron population under synchrotron, inverse
Compton and adiabatic energy losses or gains. It was demonstrated
that fossil radio plasma with an age of up to 2 Gyr can be
revived by compression in a shock wave of large-scale structure
formation, caused during the merger events or
the mass accretion of galaxy clusters. The scenario was applied to
explain the origin of the diffuse radio emission found in clusters
of galaxies, without any likely parent radio galaxy seen nearby.
The model predicts the existence of a population of diffuse
radio sources, emitting at very low 
frequencies with an ultra-steep spectrum, which are located
inside and possibly outside galaxy clusters. Thereby, extended
radio emission could be interpreted as tracing the revival of aged
fossil radio plasma which is caused by the shock waves associated with large-scale
structure formation.

Some processes related to CRs were implemented in cosmological
simulation codes. For instance, {\sl COSMOCR} is a numerical code
for the investigation of CRs in computational cosmology
\citep{2001CoPhC.141...17M}. The code includes a number of
prescriptions to account for the diffusive shock acceleration, the
mechanical and radiative energy losses and the spatial transport
of the energetic particles into the cosmic environment. Primary 
CR electrons and ions are injected at the shock sites according to the
phenomenological thermal leakage prescription. Secondary electrons
are continuously injected as a results of proton-proton (p-p) inelastic collisions
of primary CR ions and thermal background nuclei. The code
consists of a conservative, finite volume method with a power-law
sub-grid model in momentum space. Note that in this numerical
approach, the back-reaction of the non-thermal components (CRs 
and magnetic fields) caused by their pressure contribution to the thermal
gas is neglected.

To study the impact of CRs on galaxy
and cosmic structure formation and evolution, \citet{2006astro.ph..3484E}
developed an approximative framework which  treats dynamical and radiative
effects of CRs in cosmological simulations.
These authors
approximate the CR spectrum of each fluid element by a single
power-law, with spatially and temporally varying normalisation,
low-energy cut-off, and spectral index. In this framework they included
some approximate prescriptions for CR injection and
acceleration by shocks, as well as CR transport and energy losses
due to Coulomb interactions, ionisation losses, Bremsstrahlung
losses, and hadronic interactions with the background matter. These
prescriptions are suited to be included into the global schemes of
the numerical simulations of galaxy and structure formation. Although in
such implementation the description of the CR population is more
simplistic than in the work described earlier, the dynamical
influence of the CRs onto the underlying hydrodynamics is no
longer neglected. This is not only important for the dynamics of
the ICM itself but also for the injection of the CRs by shocks, which
are altered by the presence of the non-thermal pressure support of the
CRs.

\begin{figure}
\begin{center}
\includegraphics[width=0.9\textwidth]{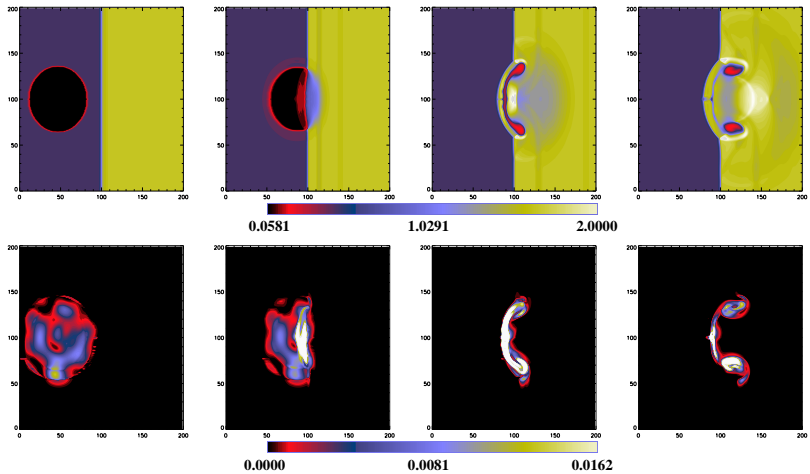}
\end{center}
\caption{Results from a simulation of a shock interacting
with a fossil radio plasma. The upper (lower) panels show the time evolution of the gas
density (magnetic field energy density) during the interaction. From \citet{2002MNRAS.331.1011E}.}
\label{fig:ens_brue_I}
\end{figure}

\subsection{The quest for radio relics}

\citet{1999ApJ...518..603R} were able to reproduce the main
features of the extended peripheral radio emission (the so-called
radio relics) observed in A~3667 by combining the single merger simulations 
(Sect.~\ref{subsec:localampl}) with a model for the in situ re-acceleration of the
relativistic particles, as described in the previous section.
 In their models, they injected, into the ICM,
relativistic electrons with a power-law spectrum, where the
power-law index $\gamma = 3/(r-1)+1$ is related to the gas
compression ratio $r=\rho_2/\rho_1$ at the shock.  They also related the age of
the radio plasma $t_{\mathrm a}$ to the distance $d=\kappa v_{\mathrm s}t_{\mathrm a}$, using a
weak-field/high-diffusion limit $\kappa=1$. By having effective
shock velocities $v_{\mathrm s}\approx 700-1000$~km\,s$^{-1}$ and
by aging the synchrotron spectrum with the formalism of
\citet{1985ApJ...291...52M}, they were able to reproduce the
observed distribution of the spectral index for a magnetic field
of $\approx 0.6\mu$G at the position of the radio relic (Fig.~\ref{fig:roet_II_a}). Since such configurations seem to be quite
common in galaxy clusters, it naturally raises the question, why not
all the clusters show such peripheral radio emission. One possible
explanation is that such shock structures are relatively short
lived compared to the merger event itself. Also, it is necessary
to have the presence of a large-scale magnetic field. It could
further be that only massive clusters can provide enough magnetic
field and strong enough merger events to trigger such a peripheral
emission.

To overcome this problem, \citet{2002MNRAS.331.1011E} proposed
that such radio relics could be made by a pre-existing fossil radio
plasma illuminated by the shock waves that the merger events originate.  In
their models, following
\citet{2001ASPC..250..454E} which take into account synchrotron,
inverse Compton and adiabatic energy losses and gains, 
they evolved the electron spectrum for the tracer
particles, which represents the fossil radio plasma.
 Their simulation, using the {\it ZEUS} code, follows the evolution of a
sphere of tracer particles hit by a shock front (Fig.~\ref{fig:ens_brue_I}).  Such a configuration nicely reproduces the
filamentary radio emission and toroidal structures observed in
many cases. These simulations also predict that  the magnetic fields are
mostly aligned with the direction of the filaments, as
suggested by observational data.

\begin{figure}
\begin{center}
\includegraphics[width=0.6\textwidth]{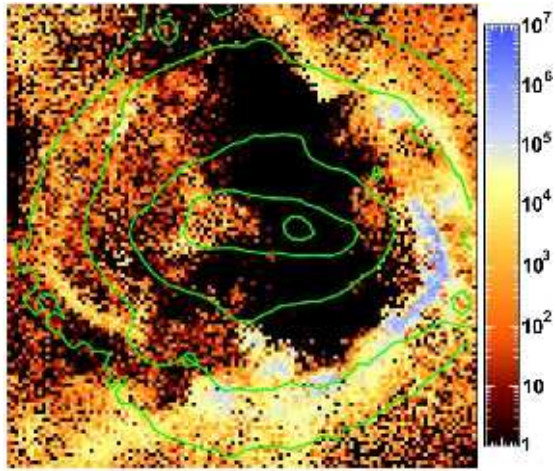}
\end{center}
\caption{Results from a simulation of  a galaxy cluster undergoing a
major merger. The figure shows the projected 'potential' radio luminosities for
1.13~Gyr old radio plasma, where $P_B/P_{\rm{gas}} = 0.01$. For
comparison, the bolometric surface X-ray luminosity (contours) is given. The
total bolometric X-ray luminosity of the cluster is
$2\times10^{44}\,\rm{erg}\,\rm{s}^{-1}$ and the emission-weighted
temperature is $3\,\rm{keV}$. Adapted from
\citet{2004MNRAS.347..389H}.} \label{fig:hoeft_I}
\end{figure}

\citet{2004MNRAS.347..389H} investigated this idea in a more realistic modelling 
by using the SPH code {\it GADGET}
\citep{SP01.1} to simulate a merging galaxy cluster within a
cosmological environment.  Such a simulation showed that the
probability for a shock wave to flare the radio plasma is highly
suppressed in the central regions of galaxy clusters, unlike 
the peripheral regions, where they found illuminated structures of
size up to Mpc scales (Fig.~\ref{fig:hoeft_I}). The
reason for this is that the radio plasma ages
much faster in the cluster centre than in the
outer regions. In fact, in the cluster centre, the pressure of the
radio plasma is higher and its energy losses due to the higher magnetic field are larger;
moreover, the
compression ratio of the shock wave is much higher in the
low-density peripheral regions than in the cluster centre. 

It is worth noting that a
necessary condition to form such relics is that the initial state
of the fossil radio plasma is characterised by a ratio of the
magnetic to thermal pressure, $P_{\mathrm B}/P_{\mathrm {gas}}$,
which has to be as low as 1 per cent to allow shocks to revive
$\approx$ 1~Gyr old radio ghosts. It is also important to mention
that \citet{2004MNRAS.347..389H} find a high probability of radio
emission, outside the shocks, due to the drained gas flows 
induced by the merger events, which transport material from the
outskirts towards the higher density regions. Therefore in some
cases the adiabatic compression seems to be sufficient to revive the
fossil radio plasma.

Recently, the picture of diffuse radio emission of galaxy
clusters has changed, as more and more detailed observations indicate
a more complex picture than the two main phenomena of radio halos and
relics (see \citealt{ferrari2008} - Chapter 6, this volume). It can well be that both scenarios described above are at
work in galaxy clusters. The in situ re-acceleration of the
relativistic particles at shocks might be responsible for the large,
external relics, whereas the compression of the fossil radio plasma might
lead to the smaller relics which are more disturbed in shape and are usually found
closer to the cluster centre.

\subsection{The quest for radio haloes}

\begin{figure}[!t]
\begin{center}
\includegraphics[width=0.48\textwidth]{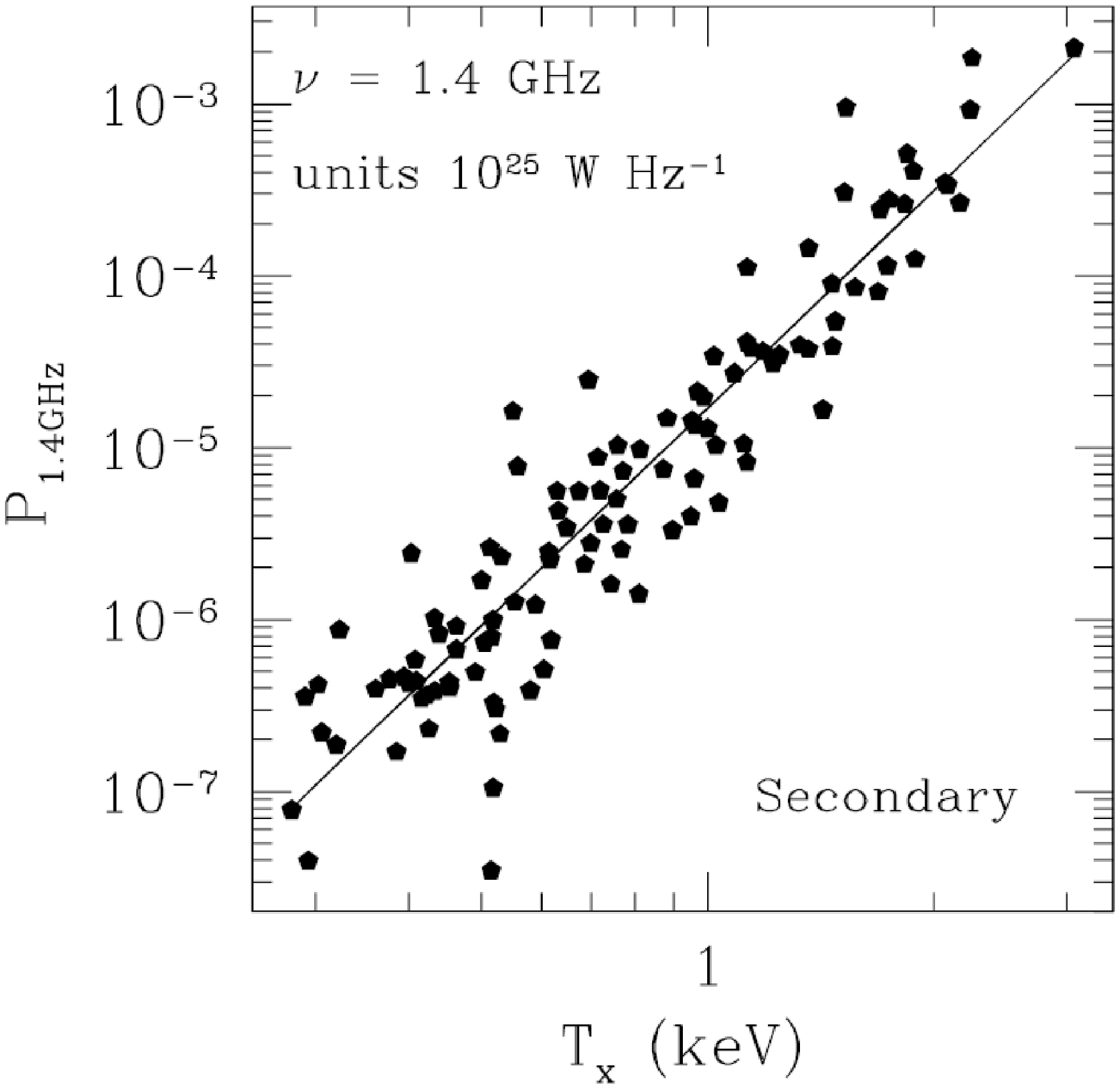}
\includegraphics[width=0.48\textwidth]{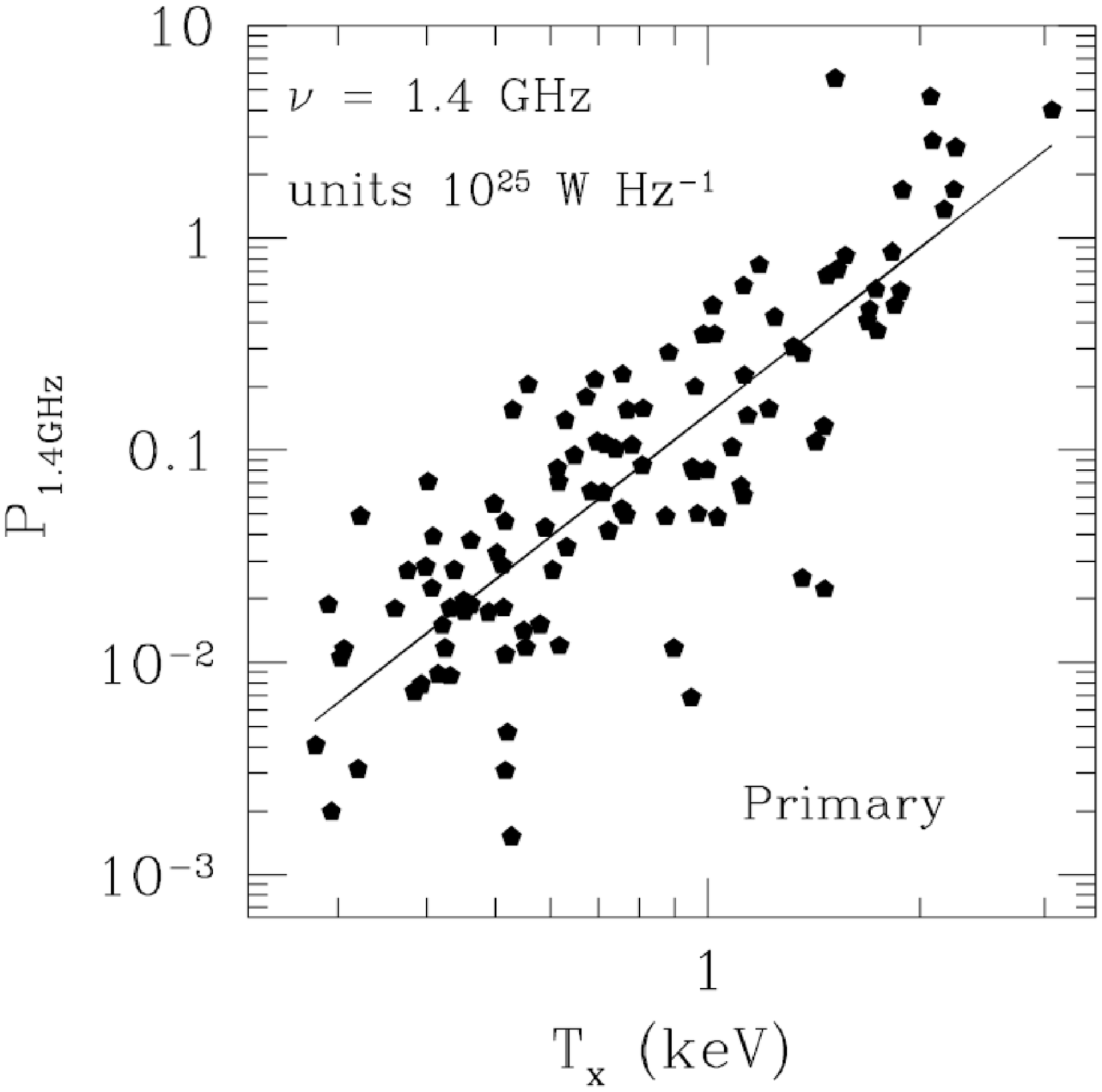}
\end{center}
\caption{Predicted radio properties of halos within a
cosmological simulation of a cubic volume with a side length of 
50 Mpc$/h$, resolved with 256$^3$ grid cells. The left panel shows
the synchrotron power of these halos at 1.4 GHz from secondary 
electrons. The right panel shows the synchrotron power from primary 
electrons, mostly located in the peripheral regions. Note that the
values for the luminosities due to the primary electrons should be scaled
with the electron to proton injection ratio $R_{{\mathrm{e/p}}}$.  From
\citet{2001ApJ...562..233M}. } \label{fig:Min_I}
\end{figure}

The accretion shocks from the LSS formation are also 
the location of the acceleration of the CRs that will be
accreted into the LSS, specially within galaxy clusters. Using the
{\it COSMOCR} code \citep{2001CoPhC.141...17M},
\citet{2001ApJ...562..233M} followed primary ions and electrons
(injected and accelerated by diffuse cosmic shocks) and secondary
electrons and positrons (produced in p-p inelastic collisions of
CR ions with thermal ICM nuclei) within a cosmological simulation.
Under the assumption that the magnetic field produced by the
battery effect reflects a fair representation of the true
distribution of the relative magnetic field strengths within the
LSS, they were able to predict the central radio emission (radio
halos), mainly produced by secondary electrons in a self-consistent
treatment (Fig.~\ref{fig:Min_I}). 

\begin{figure}[!t]
\begin{center}
\includegraphics[width=0.8\textwidth]{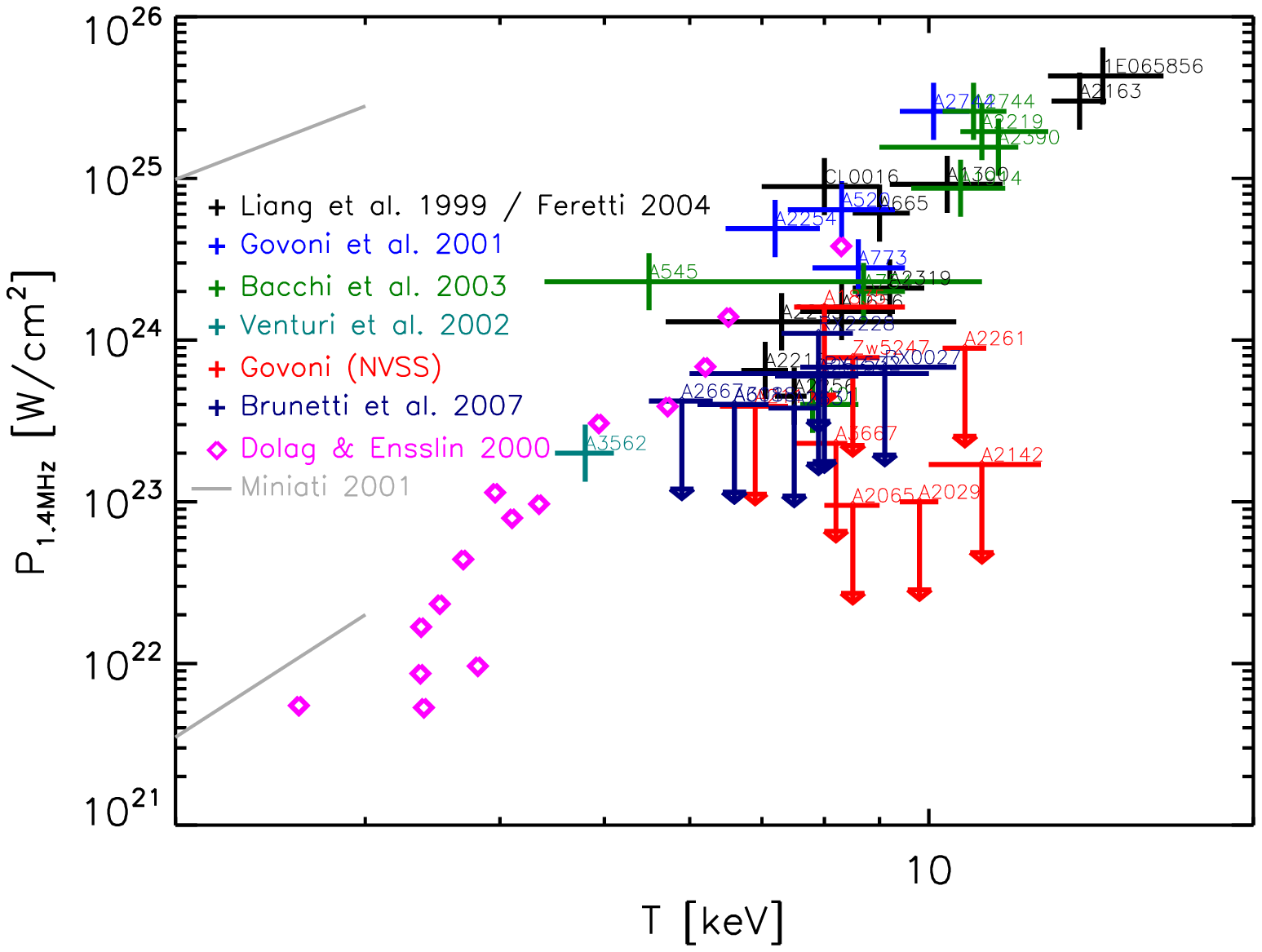}
\end{center}
\caption{Total power of radio halos observed at 1.4 MHz vs.
cluster temperature. We plot the data from
\citet{2000ApJ...544..686L}, which were partially re-observed by
Feretti (2007, in preparation) together with data from
\citet{2001A&A...369..441G,2003A&A...400..465B,2003A&A...402..913V}.
Some additional upper limits are collected thanks to the help of F.
Govoni. Additional upper limits from \citet{2007arXiv0710.0801B}
are shown. We applied a secondary hadronic model, as
described in \citet{2000A&A...362..151D},
 to calculate the radio emission from
the simulated galaxy clusters. We also added the predictions for
the emission from primary (upper line) and secondary (lower line) electrons
taken from \citet{2001ApJ...562..233M} within the temperature
range covered by the simulation. Note that the values for the
luminosities for primary electrons should be scaled with the
electron to proton injection ratio $R_{{\mathrm{e/p}}}$.} \label{fig:pt}
\end{figure}

Within their modelling they 
also reproduced the peripheral radio emission (radio relics), 
mainly produced by primary CRs, where the resulting 
morphology, polarization and
spectral index match the observed counterparts. However, one has
to note that the extrapolation of these simulations (which are on group
scale) to the observed data (which are on the scale of massive clusters) might
not be straightforward (Fig.~\ref{fig:pt}). It is further
worth commenting that even with an electron to proton injection 
ratio $R_{{\mathrm{e/p}}}$
of $10^{-2}$ (which is derived from observations, see Table~5 in
\citealt{2001ApJ...562..233M}), the predicted luminosities for the
primary emission are still significantly higher than for the
secondary one (Fig.~\ref{fig:pt}); this prediction does not seem to be
confirmed by the observations. Although these results based on
a secondary model generally agree with previous work
based on simulations of massive galaxy clusters 
\citep{2000A&A...362..151D}, they highlight the general problem that one
faces when it is applied the secondary model for relativistic electrons.
This model predicts radio haloes for every galaxy cluster, 
which is in strong contradiction with current
observations, as summarised in Fig.~\ref{fig:pt}, which also
includes the upper limits obtained by a recent observation campaign 
at the GRMT \citep[see][]{2007arXiv0710.0801B}.

\section{Turbulence}

\subsection{Simulations}

During the growth of the LSS, galaxy clusters
continuously accrete other structures, most of the time smaller
structures (like galaxies), but sometime also objects with similar
mass (major mergers). All these structures continue to move with
supersonic or transsonic velocities inside the cluster. Together
with the diffuse accretion (which is generally anisotropic) and
the generation of turbulence by hydrodynamic instabilities induced
by these bulk motions, the gas in clusters of galaxies generally
contains an amount of kinetic energy which
is not negligible compared to the amount of thermal energy.

It is worth mentioning that different simulation methods
reach good agreement in predicting that the ratio of bulk kinetic
energy to thermal energy is up to 15\% in galaxy clusters
\citep[see][]{1999ApJ...525..554F}. Based on cosmological
hydrodynamic simulations, \citet{2003AstL...29..791I} pointed out 
that the broadening of the emission lines in the X-ray band (e.g. the iron K
line) due to these expected bulk motions is appreciably
larger than the broadening due to thermal motions. Thus,
instruments like the ones designed for Suzaku or future
instruments like XEUS will be able to infer such bulk motions from
the analysis of the line shapes.

Even after subtraction of the cluster peculiar velocity,
large-scale bulk motions will be the main contributor to the
deformation of the line shapes. The imprint of turbulence induced
by hydrodynamic instabilities along flows (e.g. shear flows)
might instead be more subtle to infer from the line profiles. 

\begin{figure}
\begin{center}
\includegraphics[width=0.48\textwidth]{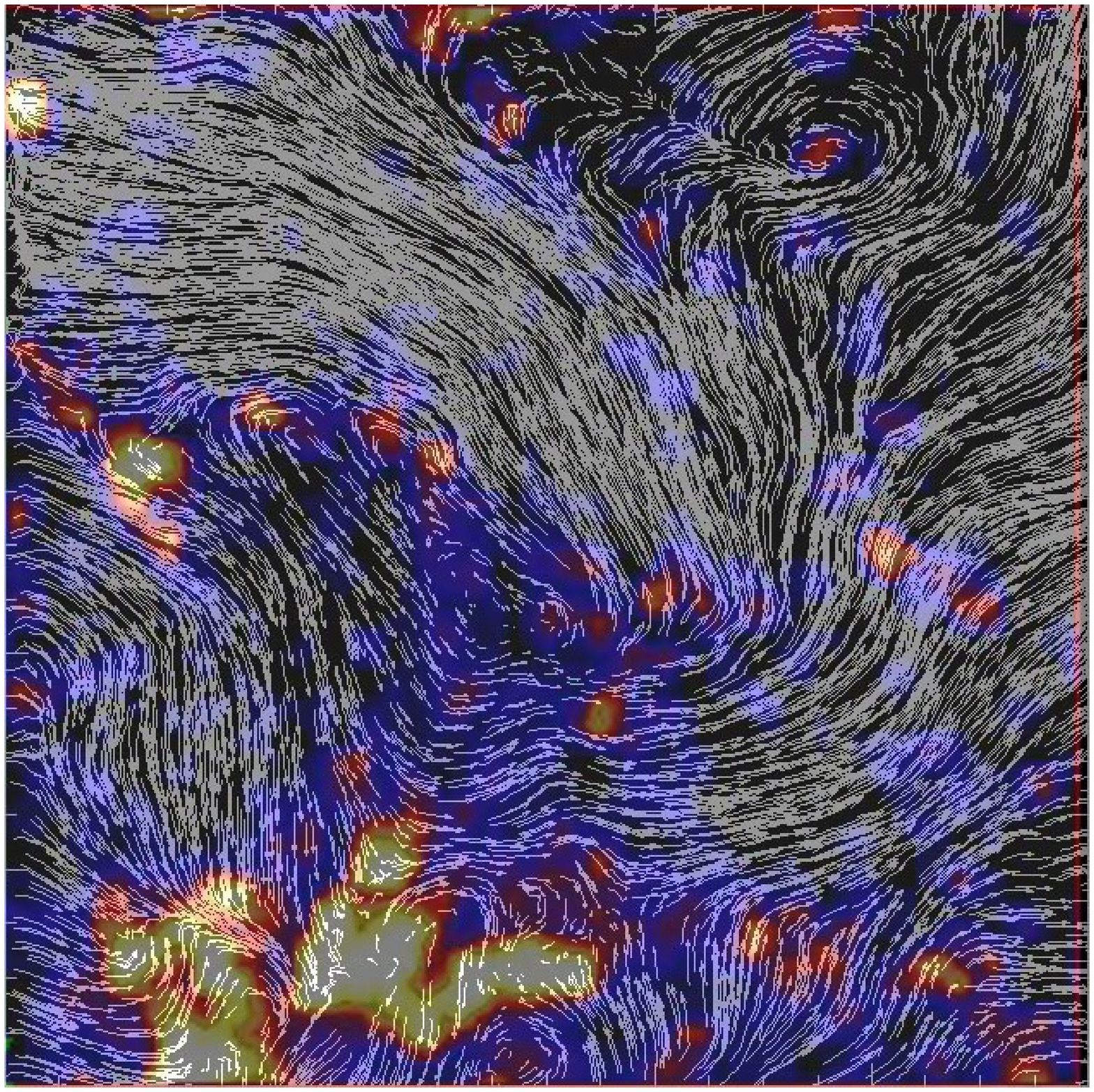}
\includegraphics[width=0.48\textwidth]{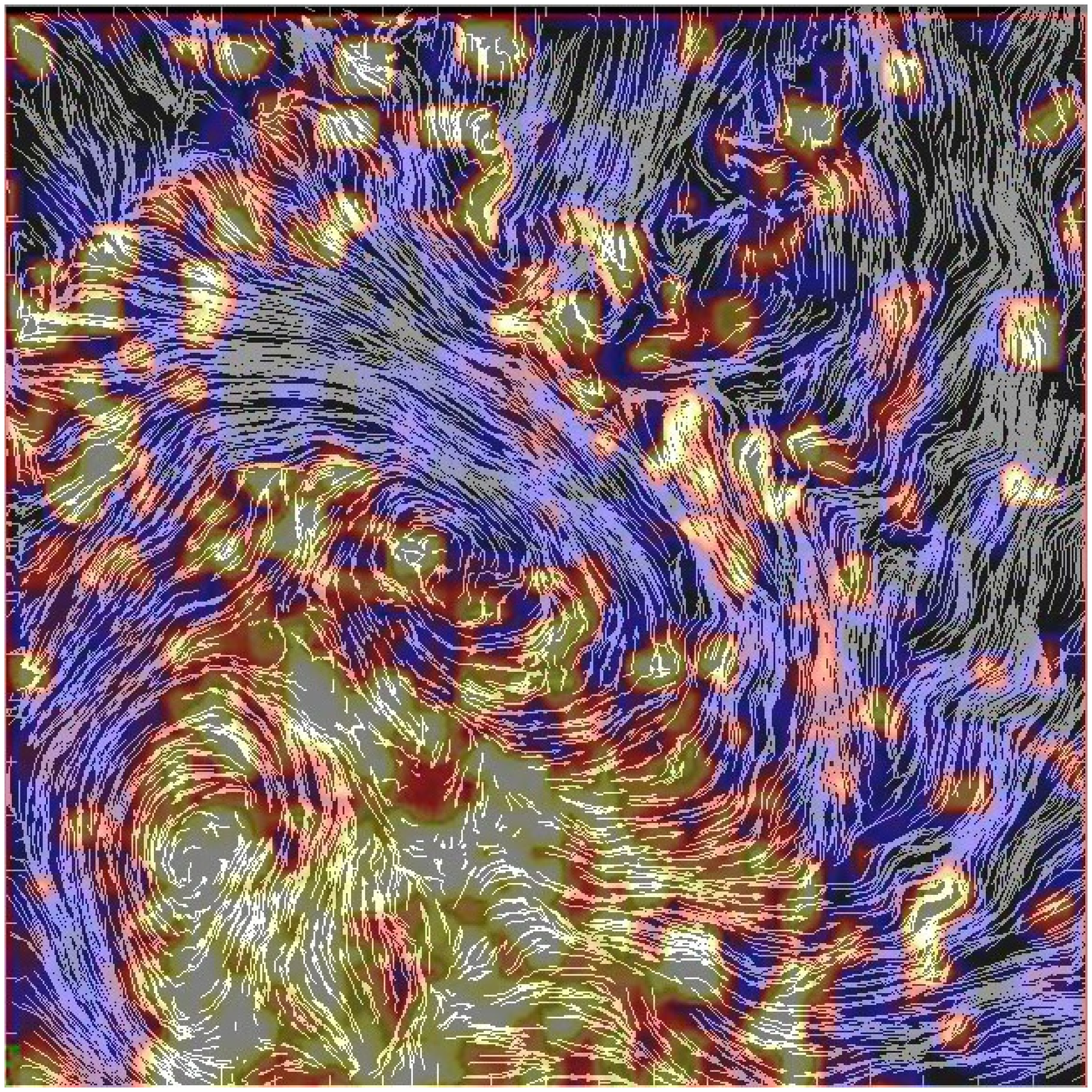}
\end{center}
\caption{The gas velocity field in a slice through the central Mpc region
of a simulated cluster after subtracting the {\em global} mean
bulk velocity of the cluster. The left panel shows a run with the
original SPH viscosity, the right panel for a low-viscosity
scheme. The underlying colour maps represent the ratio of turbulent
kinetic energy to total kinetic energy content of the particles,
inferred after subtracting the {\em local} mean velocity, as
described in \citet{2005MNRAS.364..753D}} \label{fig:dolag_turb}
\end{figure}

In recent
high-resolution SPH simulations of galaxy clusters within a
cosmological environment, \citet{2005MNRAS.364..753D} 
used a novel scheme to treat artificial
viscosity within the {\it GADGET2} code \citep{2005MNRAS.364.1105S}
to quantify how intense the shear
flow must be to drive fluid instabilities. Being a quite common process 
within cosmic structure formation, these fluid
instabilities driven by shear flows 
strongly increase the level of turbulence
within the ICM (Fig.~\ref{fig:dolag_turb}). 

Although
this small-scale turbulence contributes a measurable effect to the
non-thermal pressure within the ICM, the effect on the line shapes
is much smaller than the broadening caused by the large-scale bulk
motions. Applying this method to a set of simulated galaxy
clusters, \citet{2006MNRAS.369L..14V} showed that for relaxed
clusters the turbulent energy content in the ICM scales with the
thermal one. The fraction of turbulent to thermal energy turned 
out to be rather independent of the mass of the
system. Interestingly, this is consistent with the semi-analytical
predictions of \citet{2005MNRAS.357.1313C}, who estimated the
amount of turbulent energy within the ICM in terms of the $P{\mathrm d}V$ work
done by the infalling substructures.


\subsection{Observational evidence}

The spectroscopic resolution of current X-ray instruments is not high
enough to measure the line shapes and infer the level of
turbulence present within the ICM. However, using a mosaic of
XMM-Newton observations of the Coma cluster, 
\citet{2004A&A...426..387S} were able to produce spatially-resolved gas
pressure maps which indicate the presence of a
significant amount of turbulence. Performing a Fourier analysis of the data
reveals the presence of a scale-invariant pressure fluctuation
spectrum in the range between 40 and 90 kpc which is well
described by a projected Kolmogorov turbulence spectrum; at least
10 percent of the total ICM pressure is in
turbulent form.

Alternatively, in a more indirect way, \citet{2006A&A...453..447E}
argue that the recently reported Kolmogorov-like magnetic
turbulence spectrum, which is inferred from Faraday rotation measurements 
in the cool core of clusters, can be understood by kinetic energy
injection \citep[see also][]{2006MNRAS.366.1437S}. Such a dynamo model 
predicts the correct magnetic field
strength in cool core clusters for reasonable (but yet not
directly observable) values of the hydrodynamic turbulence velocity
and characteristic length scales. This result indicates that the magnetic
fields might directly reflect the presence of hydrodynamic
turbulence. Such models are directly related to the turbulence
induced by buoyant radio lobes from the central radio galaxy which
rise into the cluster atmosphere. On the other hand, Faraday
rotation within non-cooling flow clusters with multiple extended
radio sources (which therefore probe the magnetic field structure at
different radii) can give alternative constraints on the magnetic
field power spectra
\citep[see][]{2004A&A...424..429M,2006A&A...460..425G} 
and thus on the underlying hydrodynamic turbulence present in
the ICM.

\begin{figure}
\begin{center}
\includegraphics[width=0.5\textwidth]{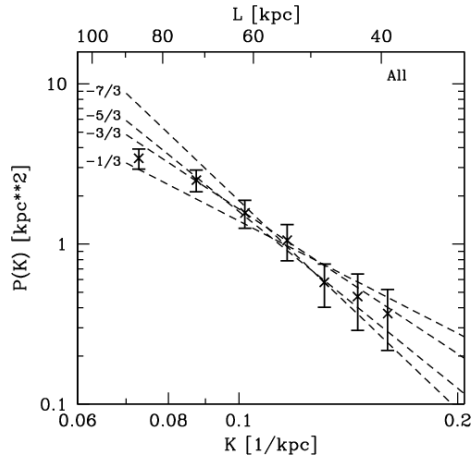}
\end{center}
\caption{The dots with error bars are the observed projected power spectral
densities as inferred from X-ray
observations of the Coma cluster, after subtraction of the shot noise; 
the dashed lines are  model predictions.
From \citet{2004A&A...426..387S}. }
\label{fig:Schuecker}
\end{figure}

\section{Concluding remarks}\label{sec:discussion}

It seems that, in the last years, a consistent picture of
magnetic fields in clusters of galaxies has emerged from both numerical
work and observations. Simulations of individual
processes like shear flows, shock/bubble interactions or
turbulence/merging events predict a super-adiabatic amplification of
magnetic fields. It is worth mentioning that this
common result is obtained by using a variety of different codes,
which are based on different numerical schemes. Further support for such
super-adiabatic amplifications comes from analytical estimates of
the anisotropic collapse which use the Zel'dovich
approximation \citep{1970A&A.....5...84Z}. When this amplification 
occurs in fully consistent cosmological simulations, various observational aspects
are reproduced (e.g. Fig.~\ref{fig:RMprof});
moreover, the final strength of the magnetic field reaches a level
sufficient to link models that predict magnetic field seed at
high redshift with the magnetic fields observed in galaxy clusters
today. It is important to note that all the simulations show
that this effect increases when the resolution is improved, and
therefore all the current values have to be taken as lower limits of
the possible amplification.

Despite this general agreement, there are significant differences 
among the 
predictions on the structure of the magnetic fields coming from different
models of magnetic field seeds. In particular, there are several
differences between the up-scaled cosmological battery fields
\citep{2001ApJ...562..233M,2004PhRvD..70d3007S} and the magnetic
field predicted by the high-resolution simulations of galaxy clusters
that use either AMR \citep{2005ApJ...631L..21B} or SPH
\citep{1999A&A...348..351D,2002A&A...387..383D,2005JCAP...01..009D}.
In the SPH case, it is possible to follow the amplification of
the field seeds within the turbulent ICM in more detail. 

A good visual
impression can be obtained by comparing the regions filled with
the high magnetic fields shown in Fig.~\ref{fig:Sigl_I} and
\ref{fig:Brueggen_I}. It is clear that the high magnetic field
regions for the battery fields are predicted to be much more
extended; this result leads to a flat profile around the forming structure,
whereas for the turbulent amplified magnetic fields the clusters
show a magnetic field distribution which is much more peaked. Part of this
difference originates from the physical model, as the
cosmological shocks are much stronger outside the clusters.
Somewhat less clear is how big it is the influence of the
different numerical resolutions of these simulations to these
discrepancies. Usually, the model parameters for such simulations
can only be calibrated using the magnetic field strength and
structure within the high-density regions of galaxy clusters.
Therefore it is crucial to perform detailed comparisons with all
the available observations to validate the simulations. Note that
extrapolating the predictions of the simulations into lower
density regions, where no strong observational constraints exist,
will further amplify the differences among the simulations in the predictions of the
magnetic field structure.

Moreover, one has to keep in mind that, depending on the ICM
resistivity, the magnetic field could suffer a decay, which is so
far neglected in all the simulations.
Furthermore, a clear deficiency of the current simulations is that they
do not include the creation of a magnetic field due to all the feedback
processes happening within the LSS (like radio bubbles inflated by
AGN, galactic winds, etc.): this might alter the magnetic field
prediction if their contribution turns out to be significant. Also
all of the simulations so far neglect radiative losses; if
included, this would lead to a significant increase of the density
in the central region of clusters and thereby to a further magnetic
field amplification in these regions.  Finally, there is an
increasing number of arguments suggesting that instabilities and
turbulence on very small scales can amplify the magnetic fields 
to the observed $\mu$G level over
a relatively short timescale.
However, it is still unclear how such fields, which are tangled on very
small scales, can be further processed to be aligned
on large scales (up to hundreds of kpc), as observed.

Concerning radio haloes and relics, the picture is only partially
consistent (for a more detailed discussion of primary and
secondary models see \citet{2004JKAS...37..493B} and references
therein as well as \citealt{ferrari2008} - Chapter 6, this volume). 
Radio relics seem to be most likely related to strong
shocks produced by major merger events and therefore produced by
direct re-acceleration of CRs, the so-called primary models.
Although some of the observed features, like morphology,
polarisation and position with respect to the cluster centre, can
be reasonably well reproduced, there might be still some puzzles
to be solved. 
On the one hand, direct acceleration of CRs in shocks
seems to overestimate the abundance, and maybe the luminosity, of
radio relics; on the other hand, simulations which illuminate
fossil radio plasma can produce reasonable relics only starting
from a small range of parameter settings. 

A similar situation
arises for modelling the central radio emission of galaxy clusters
by secondary models. The total luminosity seems
to be reproduced using reasonable assumptions and also the 
observed steep correlation between cluster temperature/mass and radio
power seems to be reproduced quite well. However, these models suffer
from two drawbacks. The first one is that, in the framework of such
models, every massive cluster produces a powerful radio halo, 
at odds with observations (Fig.~\ref{fig:pt}). 
The other problem of secondary models 
is that the detailed radio properties are not
reproduced. In fact, firstly, in most cases, the profile of radio emission 
is too steep, so that these models can almost never reproduce the size of
the observed radio halos. Secondly, the
observed spectral steepening \citep[e.g.][]{2005A&A...440..867G}
cannot be reproduced. 

One possible way out, at least for the first concern, would be
that magnetic fields are extremely dynamical and transient
and therefore {\it light up} the halo during a merger. 
However, there is also no indication from
observations that clusters which show radio emission contain 
magnetic fields which are more intense than those in clusters 
without observable extended diffuse radio emission. On the contrary, 
the cluster A~2142 has a magnetic field strength similar to the Coma cluster,
but the upper limit on its radio emission is at least two orders
of magnitude below the value expected from the correlation (Fig.~\ref{fig:pt}).  
Note that both clusters are merging systems
which are characterized by the presence of two central cD galaxies. This
indicates that there should be further processes involved or
additional conditions required to produce radio emission. It is
worth noting that recent models, based on turbulent
acceleration, seem to overcome this problem \citep[see][ and
references
therein]{2003ApJ...594..732K,2005MNRAS.363.1173B,2005MNRAS.357.1313C}.
\citet{2005MNRAS.357.1313C} predicted that the probability
for a galaxy cluster to show a giant radio halo is an increasing
function of the cluster mass; these authors also reproduced the observed probability
of $\approx 30$~\% for a massive galaxy clusters to show extended
radio emission.

Being dominated by large-scale bulk motions, induced by the
cosmological structure formation process, different simulation
methods reach good agreement in predicting that the ratio of the bulk
kinetic energy to thermal energy has an upper limit of 15~\% in galaxy
clusters \citep[see][]{1999ApJ...525..554F}. However, the exact
role of turbulence induced by hydrodynamic instabilities within
the complex flow patterns of galaxy clusters is still hard to
quantify \citep[see][]{2005MNRAS.364..753D}. It should also be
mentioned that the real viscosity of the ICM is hardly known, and
can play a noticeable role within the ICM, as demonstrated in
simulations which attempt to include physically motivated
viscosity into a cosmological context
\citep[see][]{2006MNRAS.371.1025S}.

The dynamical role of CRs in galaxy clusters is not very well
understood yet. Simulations using different codes predict quite
different relative pressure contained in CRs within galaxy
clusters. Moreover, the simulations indicate that the relative
importance of CRs in galaxy clusters also strongly depends on
other non-thermal processes, like radiative losses and feedback
from star formation \citep[see][]{2007MNRAS.378..385P}.

In general it has to be pointed out that, although the complexity of
the physical processes treated by cosmological simulations have improved
dramatically in recent years, this research field is still young.
Therefore future work is expected to enlighten such
complex processes within the ICM and will help to improve our
understanding of the non-thermal components in galaxy clusters. Such
work will also be needed to interpret the observational information
which is overwhelmingly increasing in quality and size
and is expected to become available soon with the next generation of instruments,
particularly at radio band, like ALMA, LOFAR, SKA, EVLA and others.

\section{Acknowledgments}
The authors thank ISSI (Bern) for support of the team
``Non-virialized X-ray components in clusters of galaxies''. Special
thanks to the anonymous referee for suggestions that improved the manuscript, Torsten
En{\ss}lin for various helpful discussions and 
Daniel Clarke for carefully reading the manuscript.
A.M.B. acknowledges the support from RBRF grant 06-02-16844 and a
support from RAS Programs. A.D. also gratefully acknowledges
partial support from the PRIN2006 grant ``Costituenti fondamentali
dell'Universo'' of the Italian Ministry of University and
Scientific Research and from the INFN grant PD51.

\bibliographystyle{aa}
\bibliography{15_dolag}

\end{document}